\documentclass[aip,reprint]{revtex4-1}
\usepackage{amssymb,amsmath}   
\usepackage[dvips]{graphicx}   
\usepackage{verbatim}   
\usepackage{color}      
\usepackage{subfigure}  
\usepackage{hyperref}   
\usepackage{gensymb}
\usepackage{epstopdf}
\usepackage{natbib}
\usepackage{enumerate}
\usepackage{gensymb}

\begin{document}

\title{Novel experimental design for high pressure - high temperature electrical resistance measurements in a 'Paris-Edinburgh' large volume press}
\author{Shlomi Matityahu} \email{matityas@post.bgu.ac.il}
\affiliation{Department of Physics, NRCN, P.O. Box 9001,
Beer-Sheva 84190, Israel} \affiliation{Department of Physics,
Ben-Gurion University of the Negev, Beer Sheva 84105, Israel}
\author{Moran Emuna}
\affiliation{Department of Materials Engineering, Ben-Gurion
University of the Negev, Beer Sheva 84105, Israel}
\author{Eyal Yahel}
\affiliation{Department of Physics, NRCN, P.O. Box 9001,
Beer-Sheva 84190, Israel}
\author{Guy Makov}
\affiliation{Department of Materials Engineering, Ben-Gurion
University of the Negev, Beer Sheva 84105, Israel}
\author{Yaron Greenberg}
\affiliation{Department of Physics, NRCN, P.O. Box 9001,
Beer-Sheva 84190, Israel}
\begin{abstract}
We present a novel experimental design for high sensitivity
measurements of the electrical resistance of samples at high
pressures (0-6GPa) and high temperatures (300-1000K) in a
'Paris-Edinburgh' type large volume press. Uniquely, the
electrical measurements are carried out directly on a small
sample, thus greatly increasing the sensitivity of the
measurement. The sensitivity to even minor changes in electrical
resistance can be used to clearly identify phase transitions in
material samples. Electrical resistance measurements are
relatively simple and rapid to execute and the efficacy of the
present experimental design is demonstrated by measuring the
electrical resistance of Pb, Sn and Bi across a wide domain of
temperature-pressure phase space and employing it to identify the
loci of phase transitions. Based on these results, the phase
diagrams of these elements are reconstructed to high accuracy and
found to be in excellent agreement with previous studies. In
particular, by mapping the locations of several well-studied
reference points in the phase diagram of Sn and Bi, it is
demonstrated that a standard calibration exists for the
temperature and pressure, thus eliminating the need for direct or
indirect temperature and pressure measurements. The present
technique will allow simple and accurate mapping of phase diagrams
under extreme conditions and may be of particular importance in
advancing studies of liquid state anomalies.
\end{abstract}

 \maketitle
\section{Introduction}
\label{Introduction} Physical properties of materials under
extreme conditions of pressure and temperature have been a subject
of increasing interest since the pioneering work of Bridgman in
the beginning of the last century.\cite{Bridgman} Since then, the
pressure-temperature ($P-T$) phase diagrams of a large variety of
materials, with particular emphasis on solid-solid and
solid-liquid (i.e.,\ the melting curve) phase transitions, have
been studied extensively.\cite{Young,Tonkov&Ponyatovsky} During
the last two decades, several studies have observed anomalous
behavior of structural, thermal and electrical properties in
liquids, driven by pressure and temperature. Prominent examples
are the elements phosphorus,\cite{KY00,MG03,FS04}
selenium,\cite{BVV89} sulfur,\cite{BVV91,CL05}
arsenic,\cite{TY97,CA09} bismuth,\cite{UAG92,GY09}
tellurium,\cite{BVV92,YK93,KY96} as well as the compounds
Yttrium-Alumina
[(Y$^{}_{2}$O$^{}_{3}$)$^{}_{x}$(Al$^{}_{2}$O$^{}_{3}$)$^{}_{1-x}$]\cite{AS94,GGN08,BAC09,GGN11}
and water.\cite{MO98,BI08} Such anomalies may indicate transitions
between different liquid polymorphs, distinguished by short range
atomic order.

The study of structural properties in melts, at ambient or at
elevated pressure, typically utilizes synchrotron X-ray or neutron
radiation.\cite{KY03,FA00,MM02} However, whereas the
identification of structural transitions between solid phases and
upon melting is well-developed, the state of the art for
determining structure in melts is still evolving.\cite{MM13,PFP03}
Therefore, it is desirable to develop relatively simple and
low-cost stand alone table-top measurement techniques to
complement the structural studies. There are several alternative
techniques suitable for studies of phase transitions in condensed
matter at high pressures and temperatures, such as differential
thermal analysis (DTA), thermobaric analysis (TBA), ultrasonic
measurements, and electrical resistance
measurements.\cite{Eremets} Measuring electrical and thermodynamic
properties may aid in identifying new transitions or anomalies
prior to conducting thorough structural investigations.
Furthermore, measurements of electrical and thermodynamic
properties form a complementary approach to the investigation of
phase transitions which, in conjunction with structural
measurements, provide a broad picture of the studied transitions.
In the present paper we focus on electrical resistance. This
measurement provides an efficient and sensitive probe of
structural and electronic transitions, while being relatively
simple and straightforward to implement.

Methods for electrical resistance measurements in the $P-T$ phase
space were considered in Ref.\ \onlinecite{PE08} and two pressure
cell configurations suitable for high temperatures electrical
resistance measurements using a 'Paris-Edinburgh' (PE) large
volume press,\cite{BJM92,BJM95a,BJM95b,MM99,KS04,YA11} namely
direct and indirect heating configurations, were proposed. The
direct heating configuration, meticulously developed in Ref.\
\onlinecite{PE08}, utilizes a high current directly supplied
through the anvils to the sample, and heats it. The sample is
wrapped by a cylindrical sleeve (made of graphite in Ref.\
\onlinecite{PE08}) which does not react with the sample and serves
as a pressure-transmitting medium. The electrical resistance of
the sample is extracted from the measured voltage drop on the
sample-sleeve combined assembly (generated by the high heating
current). The indirect heating configuration, on the other hand,
utilizes a high heating current supplied through the anvils to a
resistive heater. The sample, mounted in a good thermal conducting
capsule (usually BN), is insulated from the heater thereby being
indirectly heated. The electrical resistance of the sample is
measured by means of a separate (small) current, passing to the
insulated sample through two metallic electrodes. Due to the small
dimensions of the sample, combined with the cell assembly
geometry, the indirect heating configuration is more complex to
implement in the Paris-Edinburgh large volume press. However, the
indirect heating configuration has some substantial advantages,
among which are:
\begin{enumerate}[(i)]
\item Direct measurement of the potential drop across the sample
reflects its actual electrical resistance, thereby allowing higher
sensitivity to small changes in the electrical resistance of the
sample. In the direct heating technique, on the other hand, the
measured electrical resistance is the equivalent resistance of the
sample-sleeve combined assembly. If the pressure-transmitting
medium is made of a conducting material (such as graphite in Ref.\
\onlinecite{PE08}), the equivalent resistance will be less
sensitive to changes in the electrical resistance of the sample
alone.
\item Indirect heating configuration requires smaller samples
resulting in substantially smaller pressure and temperature
gradients along the sample.
\item In the direct heating configuration, abrupt changes in the
electrical resistance of the sample during a phase transition
introduce power (hence temperature) instabilities.\cite{DJD60} To
overcome this problem, a feedback algorithm involving fast data
acquisition electronic system should be introduced.\cite{PE08}
Such a problem does not exist in the indirect heating
configuration since the sample is not a part of the heating
system. Therefore, the temperature is not affected by any changes
in the electrical resistance of the sample.
\end{enumerate}

The applicability of the direct heating configuration to
electrical resistance measurements in a PE large volume press has
been demonstrated in Ref.\ \onlinecite{PE08}. We are not aware of
similar measurements in the indirect heating configuration. It is
thus advantageous to examine the feasibility of this technique and
to examine if the potential advantages relative to the direct
heating technique can be obtained in practice.

An essential ingredient of high pressure experiments is the
determination of the actual pressure and temperature conditions of
the sample. Technical issues such as high shear/tensile forces
acting on small size samples raise obstacles for a direct
measurement of the actual pressure and temperature of the sample.
Therefore, a calibration procedure relating the sample pressure
and temperature to the external, well-controlled, variables
(e.g.\, external force, oil pressure and heating power) should be
employed. It should be emphasized that different cell assemblies
and measurement protocols may posses quantitatively different
calibrations. Therefore, one should apply a calibration procedure
for any new experimental technique or a measurement protocol. The
most common calibration method, incorporated in large-scale
facilities, utilizes internal calibrants (e.g.\, NaCl, MgO, Au)
for which the equation of state is well known. Two independent
calibrants may be used for cross-calibration of pressure and
temperature.\cite{CW02,TP14} Small-scale laboratories, on the
other hand, usually derive a pressure calibration curve by
identifying well-defined phase transitions in some elements or
compounds (e.g., Bi, Tl, Ba, ZnTe), which serve as pressure
reference points.\cite{DDL72} Temperature, in many cases, is
directly measured by introducing a radial thermocouple in the
vicinity of the sample.\cite{WS09,WZ12,XL14,KY14} As opposed to
the situation in a multi-anvil high pressure apparatus, the cell
assembly in the PE press is uniaxially compressed. As a result, a
certain amount of radial flow of the pressurized cell assembly
takes place during compression. Such a flow may cause radial
displacement of the thermocouple, shifting it away from the
sample. A readout of the displaced thermocouple does not reflect
the actual temperature of the sample during the experiment. We
thus calibrate the temperature in addition to the pressure by
using several high temperature reference points in the phase
diagrams of Sn and Bi (similarly to Ref.\ \onlinecite{PE08}).

In the present paper we present a novel experimental setup for
electrical resistance measurements of materials at elevated
pressures (up to $\approx6$GPa) and temperatures (up to
$\approx1000$K), based on an indirect heating technique using a PE
large volume press. The design of such a pressure cell is
described. The high sensitivity of the present approach and its
utility in identifying phase transitions is demonstrated by
measurements on Pb, Sn and Bi. Through consideration of several
well-studied reference points in the phase diagrams of Sn and Bi,
we determine a calibration curve for the pressure and temperature
in our setup, thus eliminating the need for independent
measurements of these quantities.

The outline of the paper is as follows: In Sec.\ \ref{Sec 1} we
elaborate on the technical details of the construction of the
pressure cell. We describe the experimental setup (Sec.\ \ref{Sec
1A}) and the sample assembly (Sec.\ \ref{Sec 1B}). The system
performance is examined in Sec.\ \ref{Sec 2}. Preliminary
electrical resistance measurements of Pb, Sn and Bi samples are
presented in Sec.\ \ref{Sec 2A}. The results are discussed and
compared to previously published data. We then describe the
procedure used to calibrate the actual pressure and temperature of
the sample (Sec.\ \ref{Sec 2B}). The calibration is based on
several reference points in the phase diagrams of Sn and Bi. The
quality of the measurements and the sample pressure and
temperature calibration is tested by constructing the $P-T$ phase
diagrams of Pb, Sn and Bi and comparing them with those available
in the literature. Finally, we summarize and discuss our results
and their implications in Sec.\ \ref{Discussion}.
\section{Experimental}
\label{Sec 1}
\subsection{Experimental setup}
\label{Sec 1A} In order to control the sample pressure, we
utilized the V1-PE large volume press (piston cross
section=100$\text{cm}^2$) coupled with a computer controlled
hydraulic pump to thrust the pressing piston. The relation between
the force $F$ and the oil pressure $P^{}_{\text{oil}}$ in our
setup is $F=0.1P^{}_{\text{oil}}$, where the oil pressure is
measured in bars and the force is measured in ton-force. The oil
pressure can be increased up to about 1kbar and is monitored by a
pressure transducer. According to Refs.\ \onlinecite{BJM95a} and
\onlinecite{KS06}, a pressure gradient of 0.3GPa was found at a
pressure of 6GPa, over a distance of 3mm from the center of the
cell. Since our sample is only 0.5mm in diameter, it would be
reasonable to expect a pressure gradient of $\sim 0.05\%$ over the
entire sample.

DC power supply connected to the anvils heats the cylindrical
graphite heater (see Fig.\ \ref{fig:Our Cell} and the description
in Sec.\ \ref{Sec 1B}). Power ramping was controlled by a
programmable power module connected to the high power supply, by
which the heating power increment $\Delta W$, between subsequent
data points, is controlled. To dismiss heating rate effects on the
amplitude of the discontinuities or the corresponding values of
pressure and temperature, every experiment was repeated two to
four times with various heating rates of $3.75, 7.5, 11.25,
15$W/min.

The automated control system allows accurate control of pressure
and temperature independently. Hence, isobaric or isothermal modes
may be performed. A major drawback of the isothermal measurement
is that the sample chamber undergoes irreversible deformations
during compression and decompression. Such deformations severely
limit the number of isothermal runs which may be performed in a
given pressure cell assembly. As a result, measurements in a large
volume press are usually carried out in an isobaric mode.

At low pressures, percolation of the melt through the BN medium
(Fig.\ \ref{fig:Our Cell}) is a recurring problem when
implementing indirect heating configuration. This problem is more
pronounced during the first compression process. In order to avoid
sample percolation, we employed a protocol in which the oil
pressure is raised to $760\text{bar}$ (which is equivalent to a
sample pressure $P\approx6.3\text{GPa}$, see below), followed by a
series of isobaric measurements upon decompression. All the
results shown below were obtained using this protocol.
\subsection{Sample chamber}
\label{Sec 1B} 
\begin{figure}[ht]
\begin{center}
\subfigure[]{\label{fig:Cell_Assembly}
\includegraphics[width=0.47\textwidth,height=0.25\textheight]{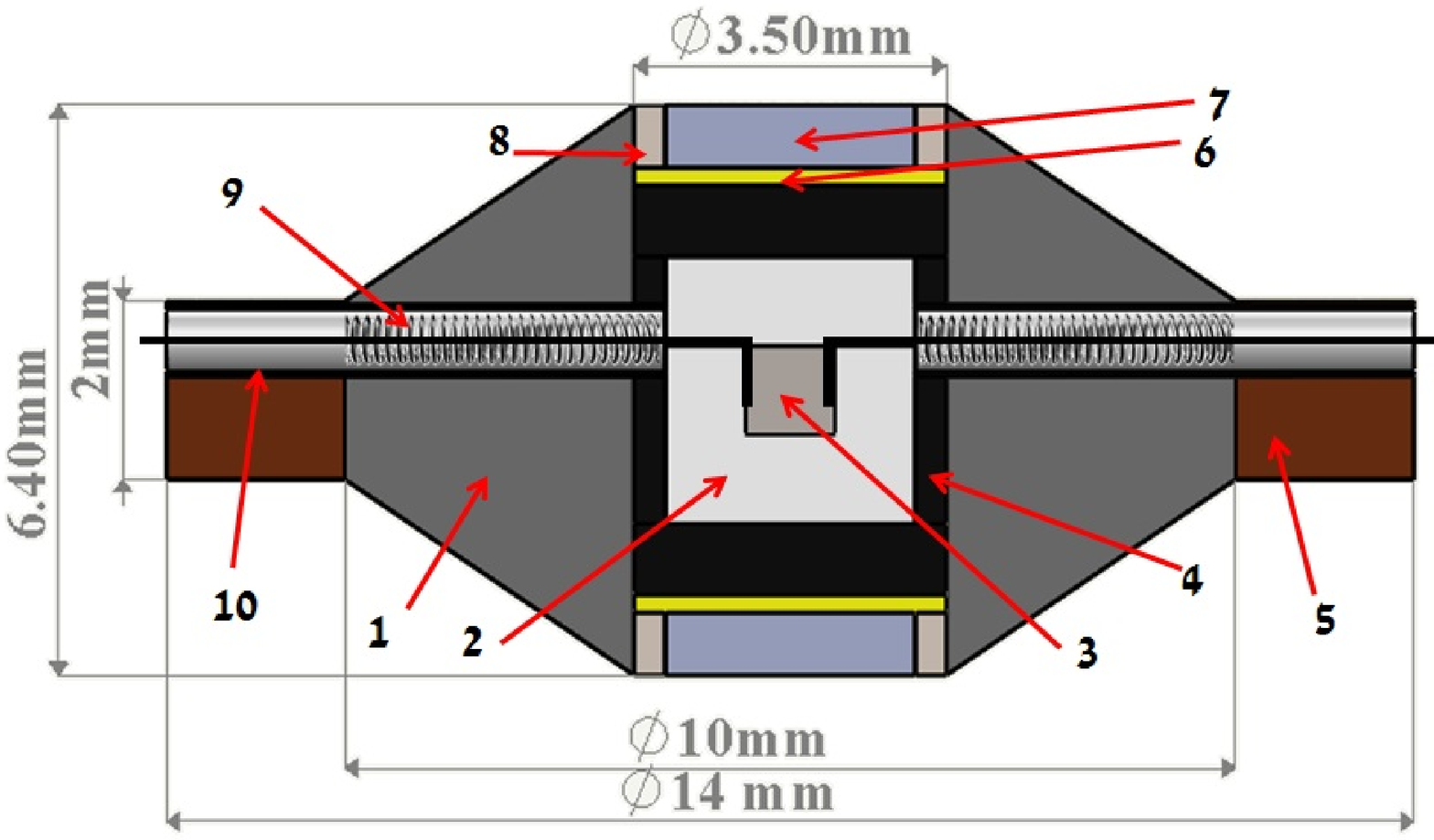}}
\subfigure[]{\label{fig:Cell_Components}
\includegraphics[width=0.35\textwidth,height=0.25\textheight]{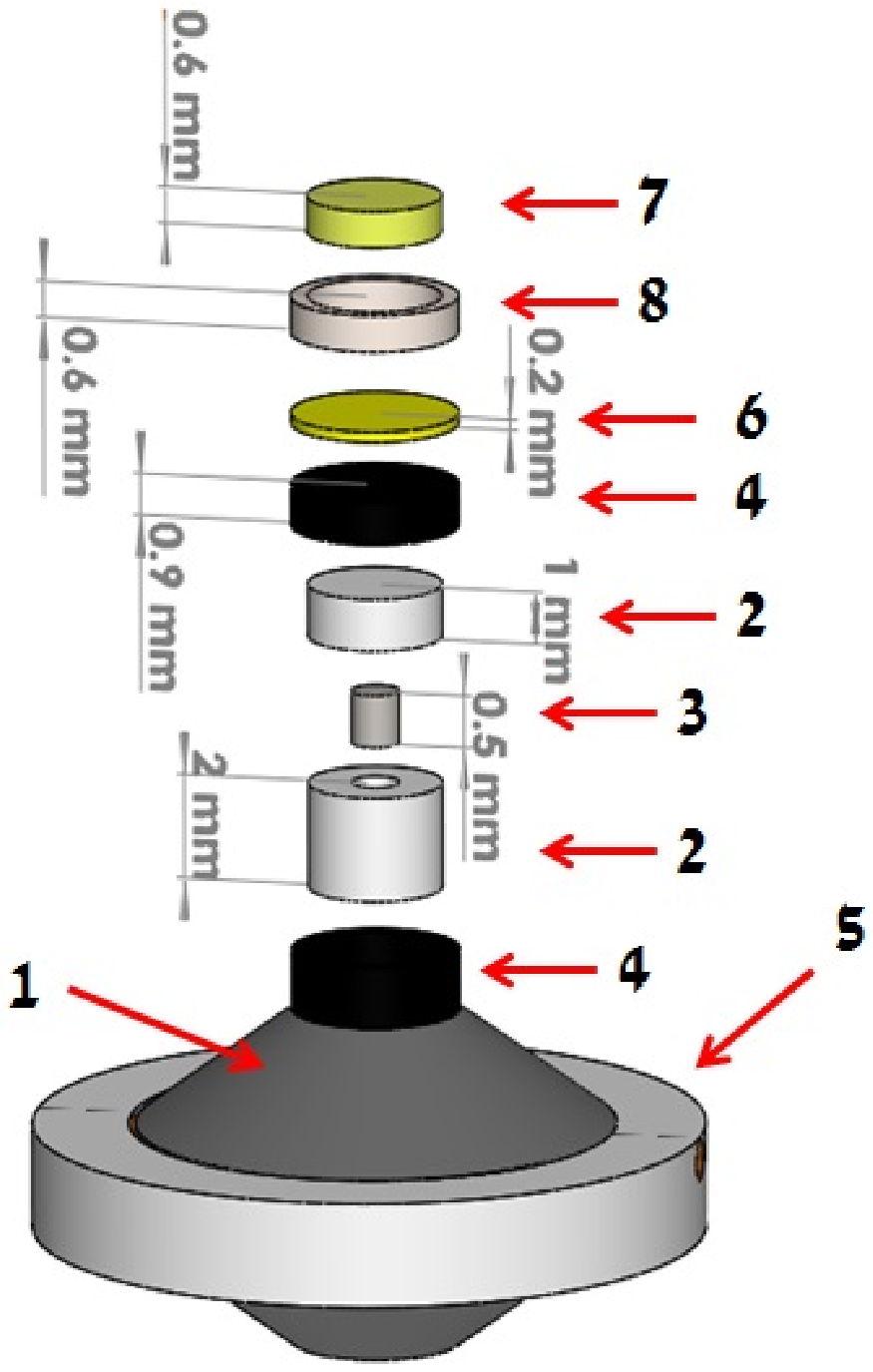}}
\end{center}
\caption{\label{fig:Our Cell}(Color online) Schematic of 10mm
bi-conical pyrophyllite cell assembly. (a) Cross-cut and (b)
"exploded" view. (1) 10mm bi-conical pyrophyllite cell (2) BN
sample chamber (3) sample (4) graphite heater (5) Teflon ring (6)
Mo disk (7) ceramic insulator (8) steel contact ring (9) coiled
tungsten electrode (10) Al$^{}_{2}$O$^{}_{3}$ sheath tube}
\end{figure}

A 10mm bi-conical pyrophyllite cell was utilized
and the cell assembly is presented in Fig.\ \ref{fig:Our Cell}. A
cylindrical sample ($0.5\times0.5$mm) was mounted in an electrical
insulating h-BN capsule (outer diameter 2.8mm). The BN capsule was
mounted within a graphite cylinder (outer diameter 3.5mm) sealed
by two graphite disks, which serves as a resistive heater. The
graphite heater is electrically connected to the anvils by means
of Mo disks and steel rings filled with MgO powder. Heating of the
graphite heater is then possible by passing a high current through
the anvils. The anvils are maintained at room temperature by
external water cooling. Two 0.2mm tungsten electrodes were
inserted to contact the sample, via radial drills through the
bi-conical pyrophyllite cell. A 0.1mm tungsten wire was coiled
around the tungsten electrodes, in order to prevent its
disconnection during compression. To avoid contact with the
graphite heater the electrical probes were sheathed by
Al$^{}_{2}$O$^{}_{3}$ tubes ($\varnothing0.8$mm). AC current, via
a well-defined shunt resistor ($R^{}_{shunt}=500\text{$\Omega$}$),
was supplied to the sample and the voltage drop across it was
recorded during the entire experiment.
\section{Results}
\label{Sec 2}
\subsection{Experimental data}
\label{Sec 2A} 
\begin{figure}[ht] \centering
\includegraphics[width=0.47\textwidth,height=0.25\textheight]{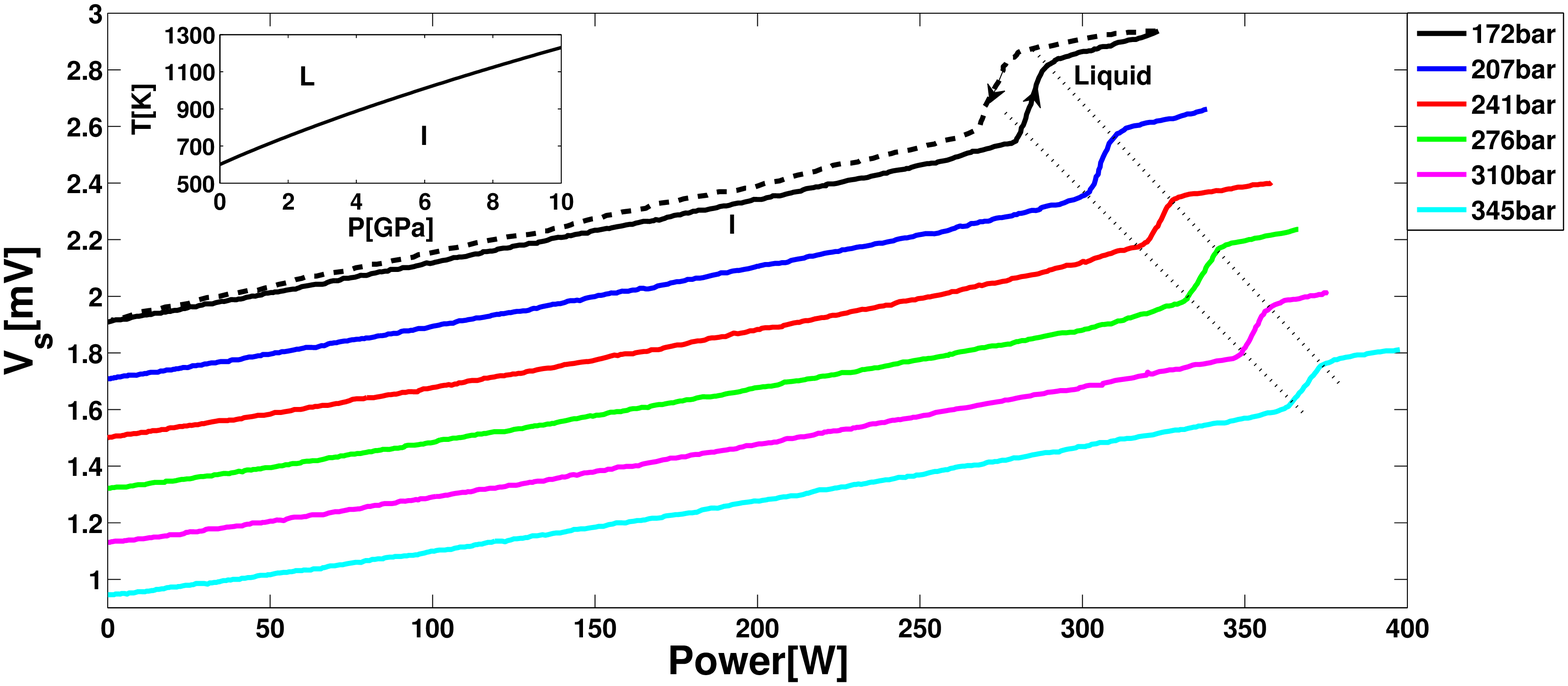}
\caption{\label{fig:Pb_Isobars}(Color online) Voltage drop across
a Pb sample as a function of heating power at selected oil
pressures. For clarity, the curves are vertically biased. The
solid (I) and liquid phases are labelled on the first isobar and
on the Pb phase diagram in the inset. For the isobar at
$P^{}_{\text{oil}}=172\text{bar}$, measurements carried out upon
heating (solid curve) and cooling (dashed curve) are presented.
Two eye-guiding dotted lines represent the beginning and ending of
the melting transition. The small distance between the lines
demonstrates the small temperature and pressure gradients across
the sample.}
\end{figure}

\begin{figure}[ht] \centering
\includegraphics[width=0.47\textwidth,height=0.25\textheight]{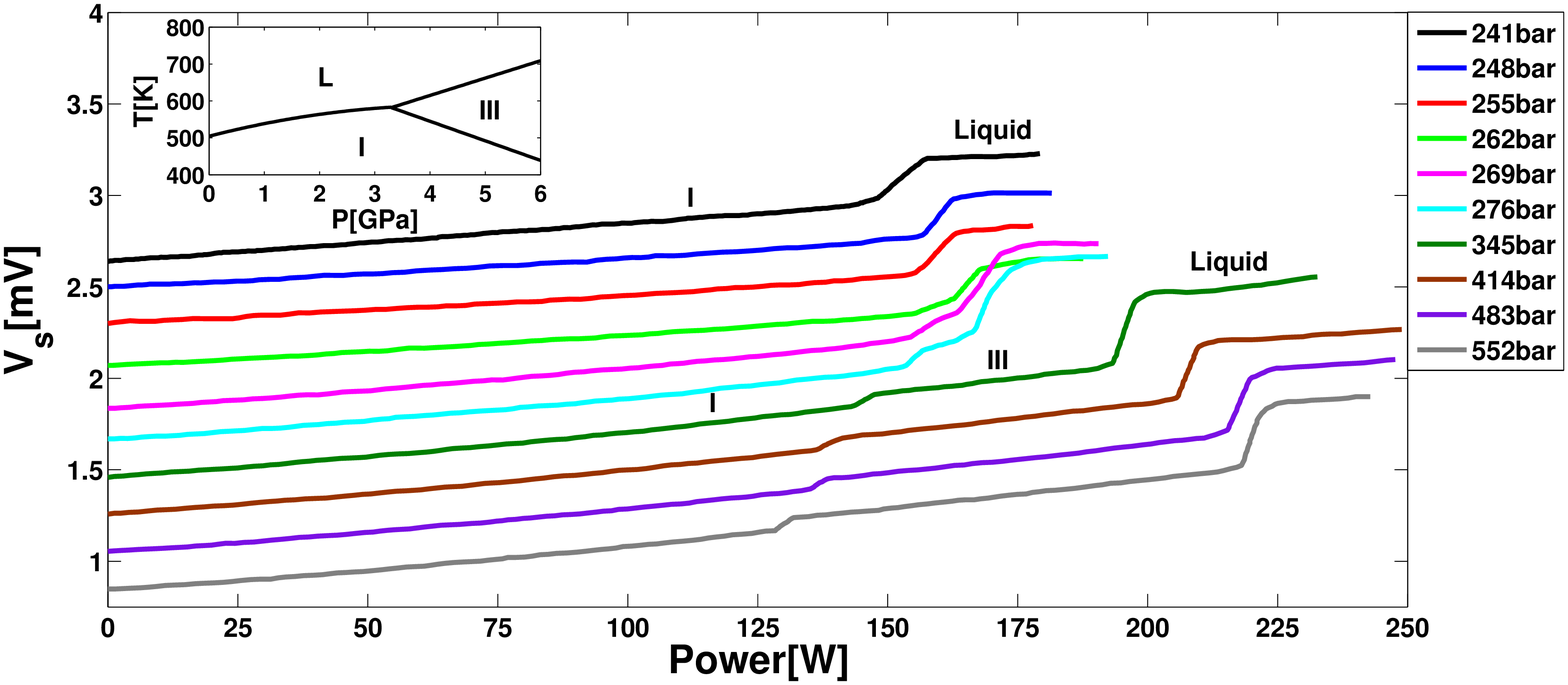}
\caption{\label{fig:Sn_Isobars}(Color online) Voltage drop across
an Sn sample as a function of heating power at selected oil
pressures. For clarity, the curves are vertically biased. The
corresponding phases are labelled on selected isobars and the Sn
phase diagram in the inset.}
\end{figure}

To validate the performance of our experimental
design, measurements were carried out on lead (Pb), tin (Sn), and
bismuth (Bi) samples. The well-known phase diagrams of these
elements exhibit increasing complexity. At moderate pressures, the
phase diagram of Pb is characterized by a single solid phase and a
melting curve exhibiting a positive slope ($dT^{}_{m}/dP>0$), as
shown in the inset of Fig.\
\ref{fig:Pb_Isobars}.\cite{Young,Tonkov&Ponyatovsky,MP76,ED10} The
melting curve of Sn is characterized by a positive slope and
includes a solid-solid-liquid triple point (see inset of Fig.\
\ref{fig:Sn_Isobars}).\cite{Young,Tonkov&Ponyatovsky,DJD60,BJD66,KAI80}
Finally, the phase diagram of Bi comprises multiple solid phases,
transitions in the liquid state,\cite{UAG92,GY09} and an anomalous
melting curve characterized by a negative slope at the low
pressures range.\cite{Young,Tonkov&Ponyatovsky} [see inset of
Fig.\ \ref{fig:Bi_Isobars1}].

\begin{figure}[ht!]
\begin{center}
\subfigure[\label{fig:Bi_Isobars1}]{\label{fig:Bi_Isobars1}\includegraphics[width=0.40\textwidth,height=0.182\textheight]{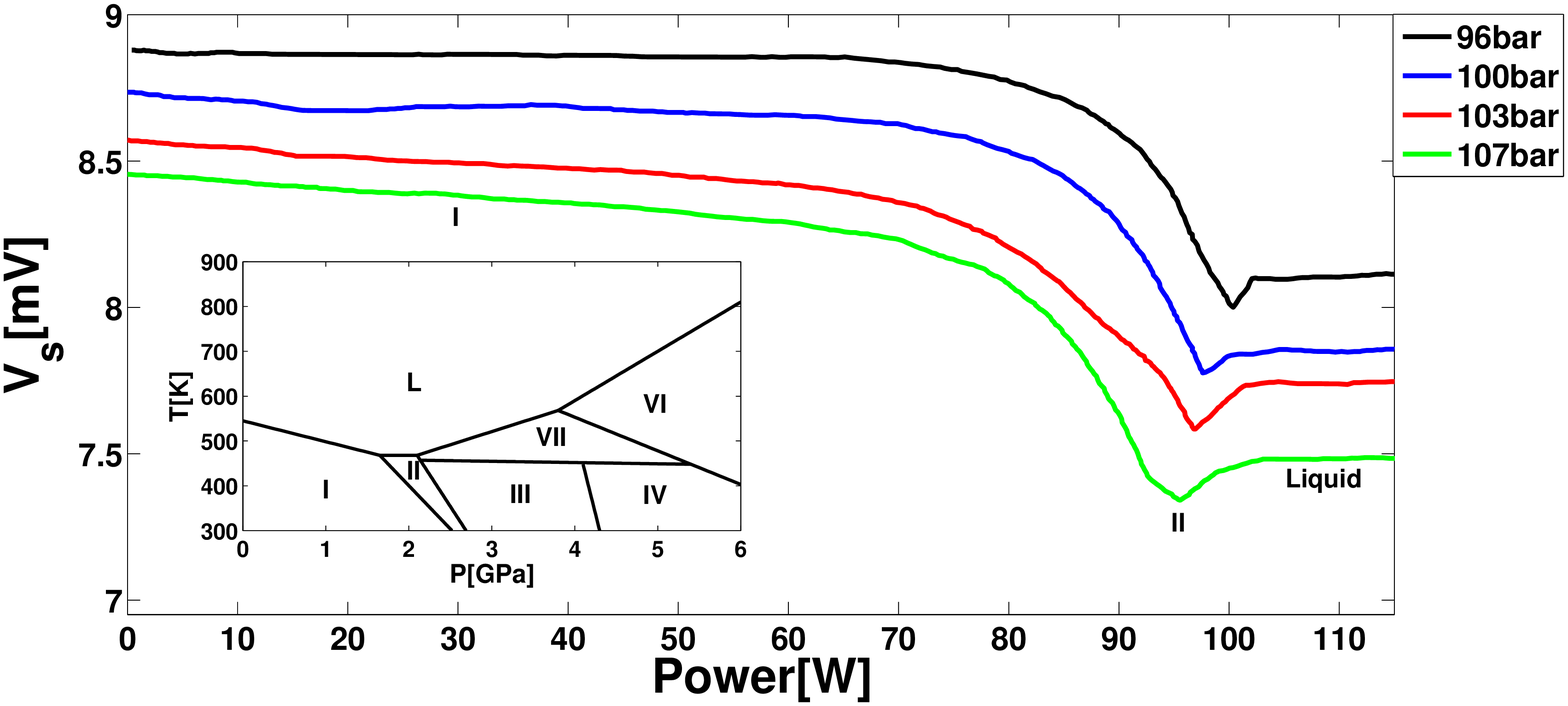}}
\subfigure[\label{fig:Bi_Isobars2}]{\label{fig:Bi_Isobars2}\includegraphics[width=0.40\textwidth,height=0.182\textheight]{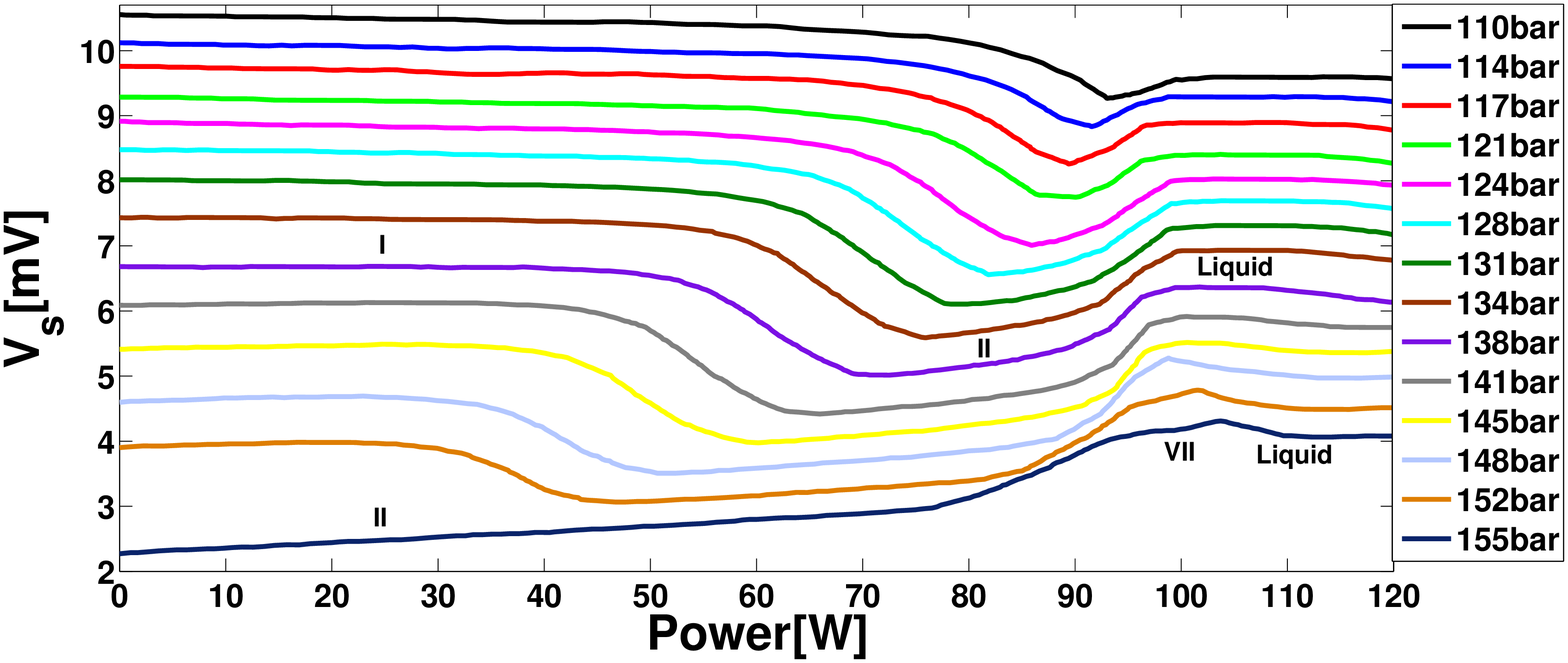}}
\subfigure[\label{fig:Bi_Isobars3}]{\label{fig:Bi_Isobars3}\includegraphics[width=0.40\textwidth,height=0.182\textheight]{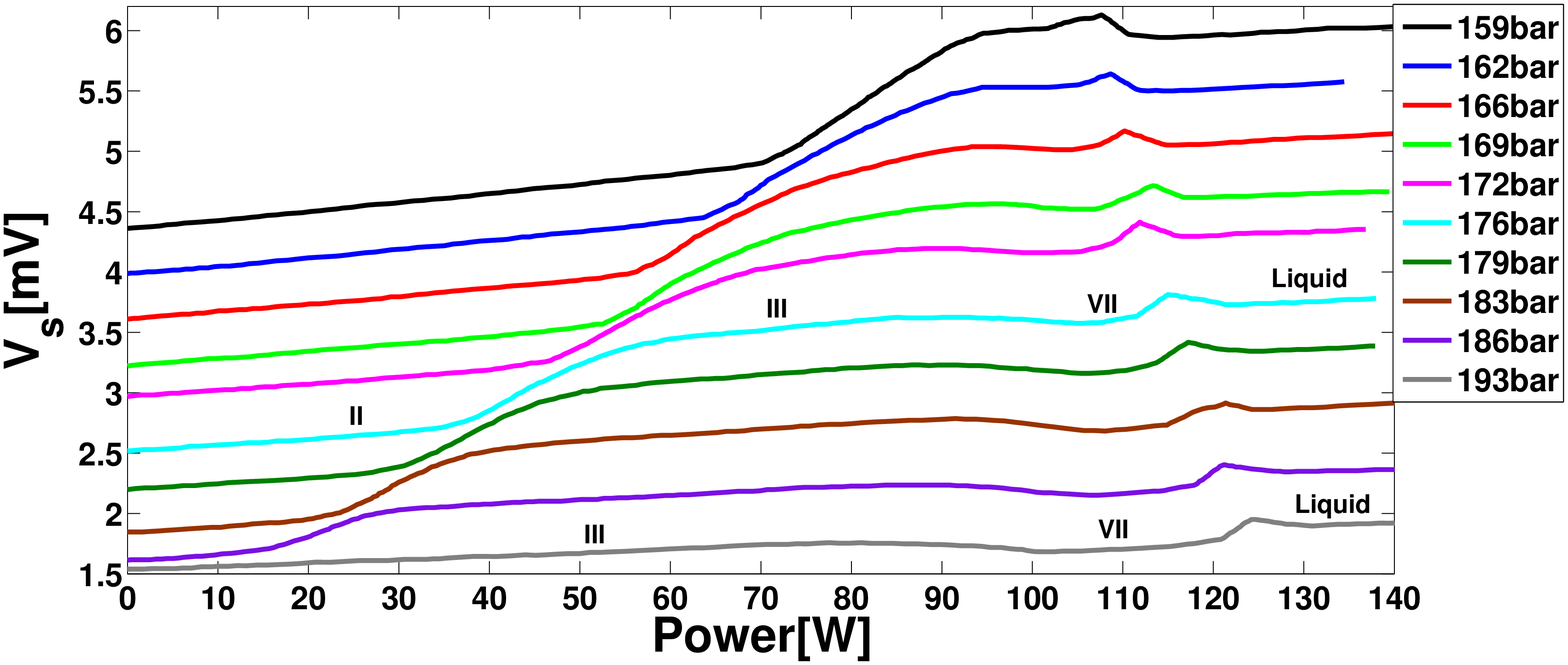}}
\subfigure[\label{fig:Bi_Isobars4}]{\label{fig:Bi_Isobars4}\includegraphics[width=0.40\textwidth,height=0.182\textheight]{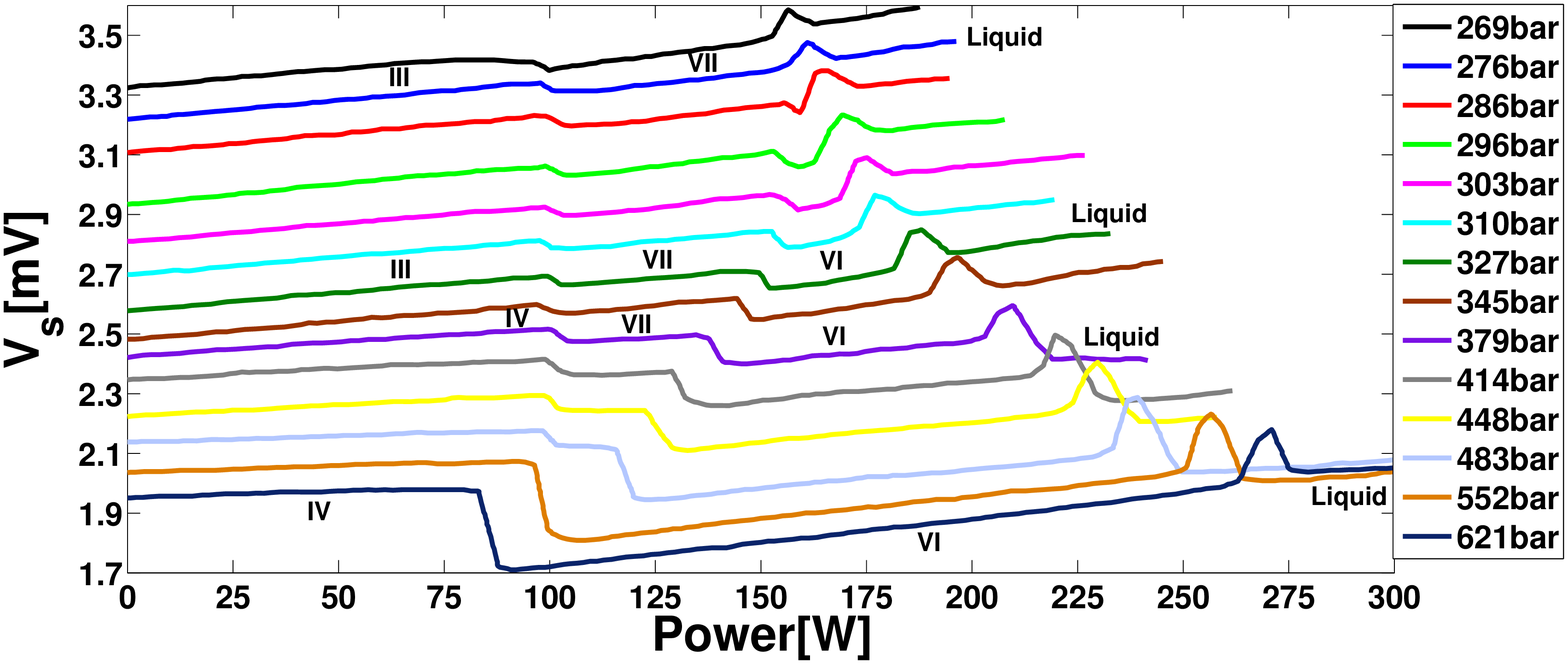}}
\end{center}
\caption{\label{fig:Bi_Isobars}(Color online) Voltage drop across
a Bi sample as a function of heating power at selected oil
pressures. For clarity, the curves are vertically biased. The
corresponding phases are labelled on selected isobars and on the
Bi phase diagram in the inset of (a).}
\end{figure}

Figures \ref{fig:Pb_Isobars}, \ref{fig:Sn_Isobars} and
\ref{fig:Bi_Isobars} show a series of isobaric measurements of the
voltage drop (proportional to the electrical resistance of the
sample) across the sample as a function of heating power, measured
upon heating of Pb, Sn and Bi samples, respectively. The data
obtained upon cooling exhibit the same behavior, apart from an
hysteresis characteristic of first-order phase transitions, as
shown for the first isobar in Fig.\ \ref{fig:Pb_Isobars}.

The measurements performed on Pb (Fig.\ \ref{fig:Pb_Isobars})
exhibit a sharp "step-like" discontinuity which corresponds to a
solid-liquid phase transition (melting). The transition
temperature increases with pressure, as expected from the positive
slope of the melting curve of
Pb.\cite{Young,Tonkov&Ponyatovsky,MP76,ED10} At low pressures, the
Sn measurements (Fig.\ \ref{fig:Sn_Isobars}) exhibit a similar
trend as Pb. However, an additional phase transition is present at
$P^{}_{\text{oil}}\approx 262\text{bar}$, accompanied by a smaller
discontinuity in the electrical resistance. This point is
attributed to the I-III-L triple point in the phase diagram of Sn
(see inset of Fig.\ \ref{fig:Sn_Isobars}). For
$P^{}_{\text{oil}}>262\text{bar}$, the temperature range in which
phase III exists enlarges with increasing pressure.
Simultaneously, the existence range of phase I decreases.

Results of the electrical resistance measurements of Bi are
presented in Fig.\ \ref{fig:Bi_Isobars}. The measurements shown in
Fig.\ \ref{fig:Bi_Isobars} are in excellent agreement with
previous electrical resistance measurements.\cite{BFP58} The
transition from a semi-metallic phase (phase I) to a metal (phase
II) is well-studied \cite{BPW35,BPW52} and is widely used for
pressure calibration. This transition is characterized by a sharp
drop in electrical resistance.\cite{BPW52,BFP58} Such a sharp drop
can be clearly observed in Figs. \ref{fig:Bi_Isobars1} and
\ref{fig:Bi_Isobars2}. As the pressure increases, this transition
occurs at higher heating powers. This implies that the I-II
coexistence curve in the $P-T$ phase space is characterized by a
negative slope, $dT/dP<0$. The II-L (melting) transition
temperature, on the other hand, exhibits almost no pressure
dependence. These findings are in qualitative agreement with the
Bi phase diagram [see inset of Fig.\ \ref{fig:Bi_Isobars1}].

As the pressure increases, the stability range of phase II expands
while that of phase I shrinks [see Figs.\ \ref{fig:Bi_Isobars1}
and \ref{fig:Bi_Isobars2}]. At $P^{}_{\text{oil}}\approx
155\text{bar}$ and above, phase I does not exist above room
temperature. The triple point II-VII-L can also be identified from
the measurements shown in Fig.\ \ref{fig:Bi_Isobars2}. It is
readily seen that at $P^{}_{\text{oil}}\approx 148\text{bar}$ an
additional phase begins to develop before the melting transition,
with the transition accompanied by a decrease in the electrical
resistance. This discontinuity is associated with the II-VII
transition. The range of phase VII increases with increasing
pressure, as seen in Fig.\ \ref{fig:Bi_Isobars3}, along with the
appearance of the II-III transition, accompanied by a large
increase in the electrical resistance. With increasing pressure,
phase II disappears in favor of phase III and the transition
temperature strongly decreases with increasing pressure. At
$P^{}_{\text{oil}}\approx 193\text{bar}$ phase II does not exist
above room temperature. At $P^{}_{\text{oil}}\approx
286\text{bar}$ an additional transition appears before the melting
transition, accompanied by a further sharp decrease in the
electrical resistance (see Figure \ref{fig:Bi_Isobars4}). This
point corresponds to the VI-VII-L triple point. As pressure is
further increases, the stability range of phase VI expands while
that of phase VII shrinks. The III-VII transition temperature,
however, does not shift with increasing pressure. At
$P^{}_{\text{oil}}\approx 552\text{bar}$ phase VII is not stable
anymore, which indicates the existence of a IV-VI-VII triple
point. It should also be noted that the melting curve of phases
VII and VI [Figs.\ \ref{fig:Bi_Isobars3} and
\ref{fig:Bi_Isobars4}] is characterized by a positive slope
($dT^{}_{m}/dP>0$). All these observations are in line with the Bi
phase diagram [inset of Fig.\ \ref{fig:Bi_Isobars1}].

As mentioned above, measurements were carried out isobarically.
Such a thermodynamic path may prevent the detection of the almost
vertical coexistence curve of phases III and IV. However, by
careful inspection of Fig.\ \ref{fig:Bi_Isobars4} one may observe
a change in the slope $dV^{}_{s}/dW$ of the low-pressure phase at
$P^{}_{\text{oil}}\approx 379\text{bar}$; the slope $dV^{}_{s}/dW$
of the low-pressure phase for $P^{}_{\text{oil}}\leq
345\text{bar}$ is slightly larger than that of the low-pressure
phase for $P^{}_{\text{oil}}\geq 379\text{bar}$. This observation
suggests that the phase transition III-IV occurs at room
temperature for $345\text{bar}<P^{}_{\text{oil}}<379\text{bar}$.

All measurements discussed above incorporate various solid-solid
and solid-liquid phase transitions readily identified by clear and
sharp discontinuities. These measurements were found to be
reproducible upon several heating and cooling cycles. In
comparison to measurements in Sn and Bi samples using the direct
heating configuration, employed in Ref.\ \onlinecite{PE08}, the
following difference should be emphasized. In the direct heating
configuration, phase transitions in Sn and Bi are identified by
changes in the slope of the electrical resistance as a function of
heating power. In many transitions the change in the slope is
rather small, making it difficult to observe phase transitions
unambiguously. Two possible sources may contribute to the smearing
of the sharp transitions in the direct heating configuration: (i)
The electrical resistance of the graphite pressure-transmitting
medium which serves as a parasitic resistor connected in parallel
to the sample, hence diminishes the sample role in the measured
equivalent resistance. (ii) The significant pressure and
(especially) temperature gradients across the sample length in the
direct heating configuration. These result in a gradient in the
electrical resistance of the sample, which smears the sharp
discontinuities at phase transitions.

Both of these problems are absent in the indirect heating
configuration. The measured electrical resistance is that of the
sample itself and the small sample dimensions limit the pressure
and temperature gradients. In the next section we use the above
measurements to calibrate the sample pressure and temperature in
terms of the oil pressure and heating power. Based on this
calibration, we then derive the $P-T$ phase diagrams of Pb, Sn and
Bi.
\subsection{Pressure and temperature Calibration}
\label{Sec 2B} As discussed in the introduction, we apply a simple
fitting procedure to several reference points in the phase
diagrams of Sn and Bi. These reference points are labelled in the
insets of Fig.\ \ref{fig:Calibration a} and discussed in Sec.\
\ref{Sec 2A}. To relate the sample pressure and temperature to the
oil pressure ($P^{}_{\text{oil}}$) and heating power ($W$), we
assume the following relations:\cite{PE08}
\begin{align}
\label{eq:1}&P=P^{}_{\infty}\left(1-e^{-P^{}_{\text{oil}}/P^{}_{0}}\right),\\
\label{eq:2}&T=a(P)\cdot W+T^{}_{0}.
\end{align}
Equation \eqref{eq:1} consists of two fitting parameters,
$P^{}_{\infty}$ and $P^{}_{0}$, which are assumed to be
temperature independent. As in Ref.\ \onlinecite{PE08}, we found
that this assumption remains valid for temperatures up to 1000K,
the maximum temperature explored in our experiments. The
exponential decay function in Eq.\ \eqref{eq:1} reflects the
gasket deformation and accounts for the deviations from linear
response at higher pressures.\cite{PE08} At a given pressure, Eq.\
\eqref{eq:2} assumes a linear variation of the sample temperature
as a function of heating power, as verified in previous
works.\cite{PE08,KY14} The slope of the straight line was shown to
be slightly pressure dependent.\cite{PE08,KY14} We thus assume
$a(P)=a^{}_{0}+a^{}_{1}P$, where $a^{}_{0}$ and $a^{}_{1}$ are
parameters to be fitted. The parameter $T^{}_{0}$ in Eq.\
\eqref{eq:2} describes the sample temperature in the absence of
heating power and is taken as $T^{}_{0}=295\text{K}$. Variations
of $\pm10\text{K}$ in this value produce very small changes in the
values of the fitting parameters $a^{}_{0}$ and $a^{}_{1}$.

Following Ref.\ \onlinecite{PE08}, we first calibrate the sample
pressure using six reference points in the phase diagram of Bi and
one triple point in the phase diagram of Sn, as labelled in the
inset of Fig.\ \ref{fig:Calibration a}. The fitting parameters are
found to be $P^{}_{0}=414\text{bar}$ and
$P^{}_{\infty}=7.5\text{GPa}$ and the corresponding calibration
curve is shown in Fig.\ \ref{fig:Calibration a}. The uncertainty
in the sample pressure is estimated to be $5\%$. Having calibrated
the sample pressure, we used a two-dimensional fitting procedure
to fit the sample temperature according to Eq.\ \eqref{eq:2}. The
fitting parameters are found to be $a^{}_{0}=1.82\text{KW}^{-1}$
and $a^{}_{1}=-0.027\text{KW}^{-1}\text{GPa}^{-1}$ and the
corresponding calibration surface, $T=T(P,W)$, is shown in Fig.\
\ref{fig:Calibration b}. The uncertainty in the sample temperature
is estimated to be 20K.
\begin{figure}[ht]
\begin{center}
\subfigure[\label{fig:Calibration a}]{ \label{fig:Calibration a}
\includegraphics[width=0.47\textwidth,height=0.25\textheight]{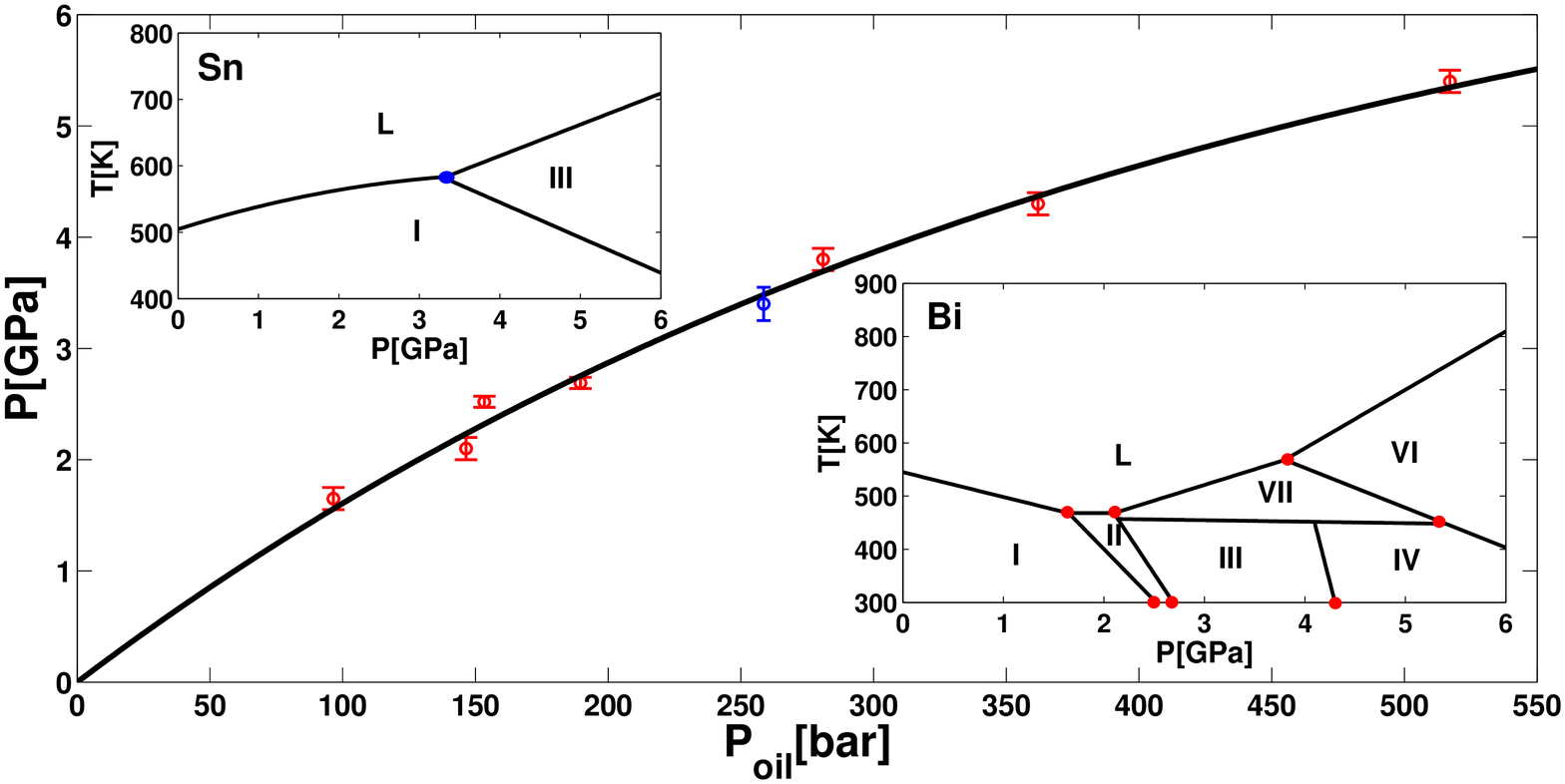}}
\subfigure[\label{fig:Calibration b}]{ \label{fig:Calibration b}
\includegraphics[width=0.44\textwidth,height=0.25\textheight]{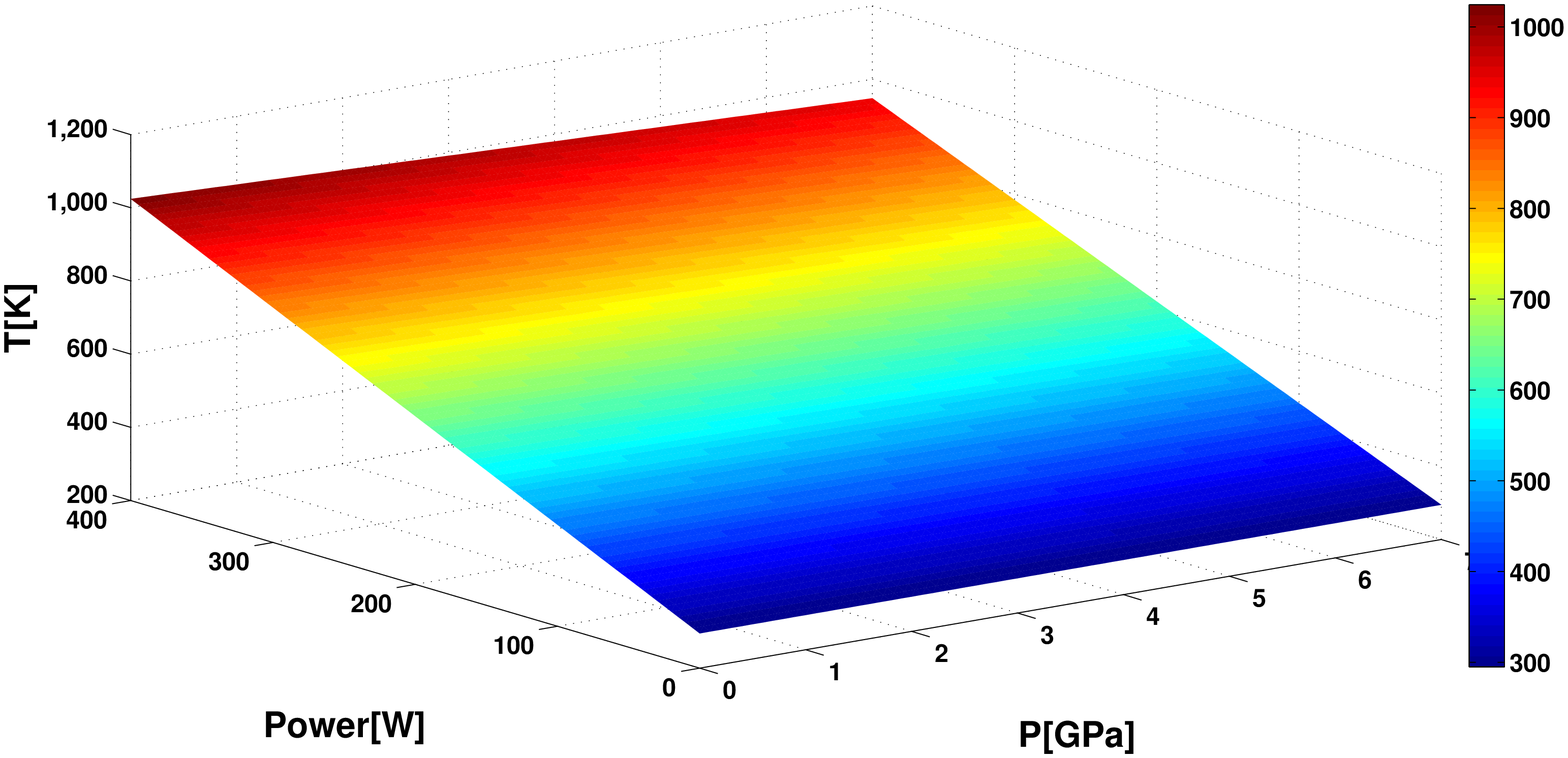}}
\end{center}
\caption{\label{fig:Calibration}(Color online) (a) Sample pressure
calibration curve, $P=P(P^{}_{\text{oil}})$, derived from a
fitting to Eq.\ \eqref{eq:1}, based on several reference points in
the phase diagrams of Sn and Bi as labelled in the insets. (b)
Sample temperature calibration surface, $T=T(P,W)$, derived from a
fitting to Eq.\ \eqref{eq:2}, based on the high-temperature
reference points also used for pressure calibration.}
\end{figure}

It should be stressed that the sample pressure and temperature
calibration in Fig.\ \ref{fig:Calibration} applies only for the
protocol used in this work, i.e.,\ for a compression to
$P^{}_{\text{oil}}=760\text{bar}$ followed by a decompression
process. Indeed, the values of the fitting parameters $P^{}_{0}$,
$P^{}_{\infty}$, $a^{}_{0}$ and $a^{}_{1}$ are quite different
from those evaluated in Ref.\ \onlinecite{PE08}. This may be
attributed to inherent differences between the direct and indirect
heating configurations, but also to the different protocol used in
Ref.\ \onlinecite{PE08}, in which measurements were performed upon
compression.

\begin{figure}[!htbp]
\begin{center}
\subfigure[\label{fig:Phase Diagrams a}]{ \label{fig:Phase
Diagrams
a}\includegraphics[width=0.47\textwidth,height=0.25\textheight]{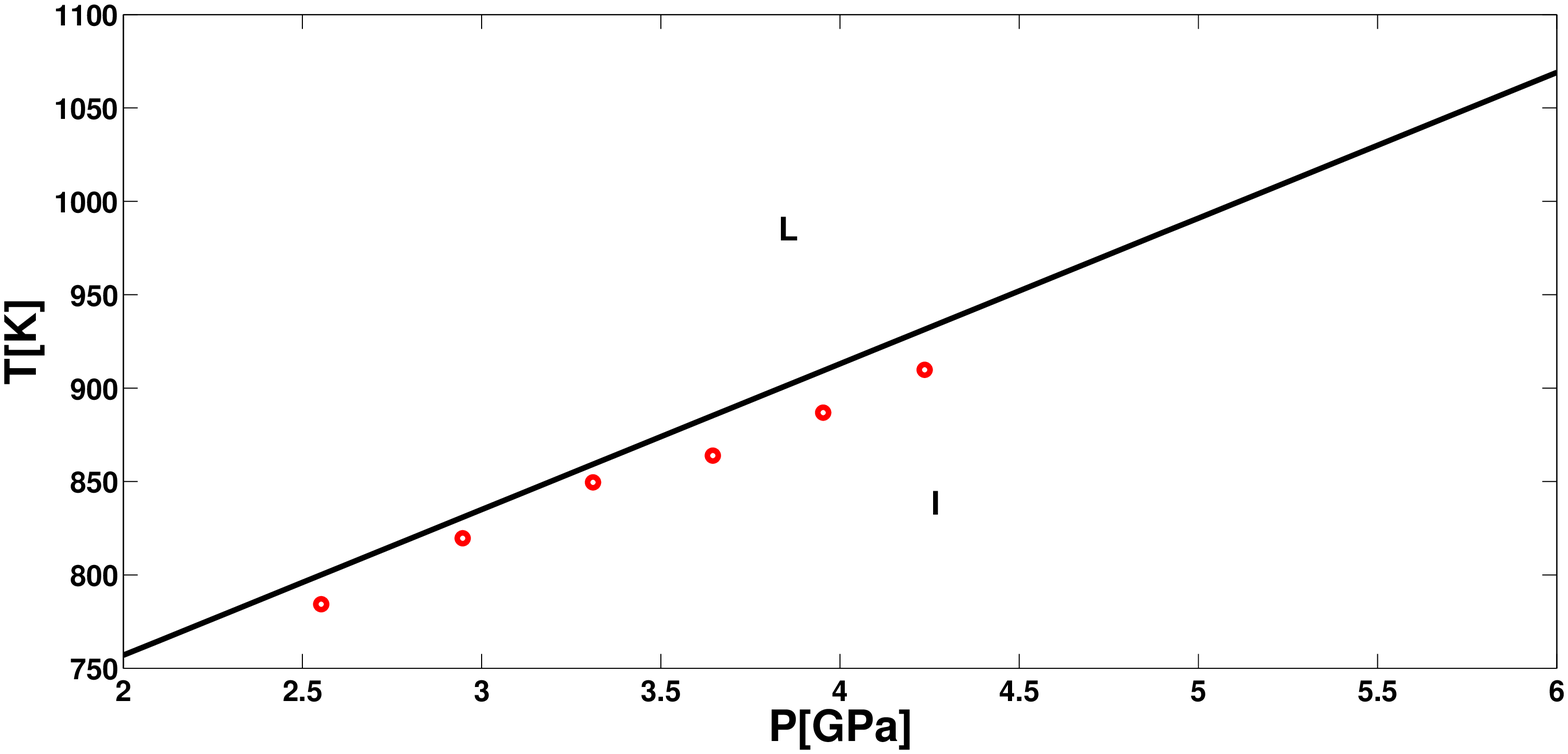}}
\subfigure[\label{fig:Phase Diagrams b}]{ \label{fig:Phase
Diagrams
b}\includegraphics[width=0.47\textwidth,height=0.25\textheight]{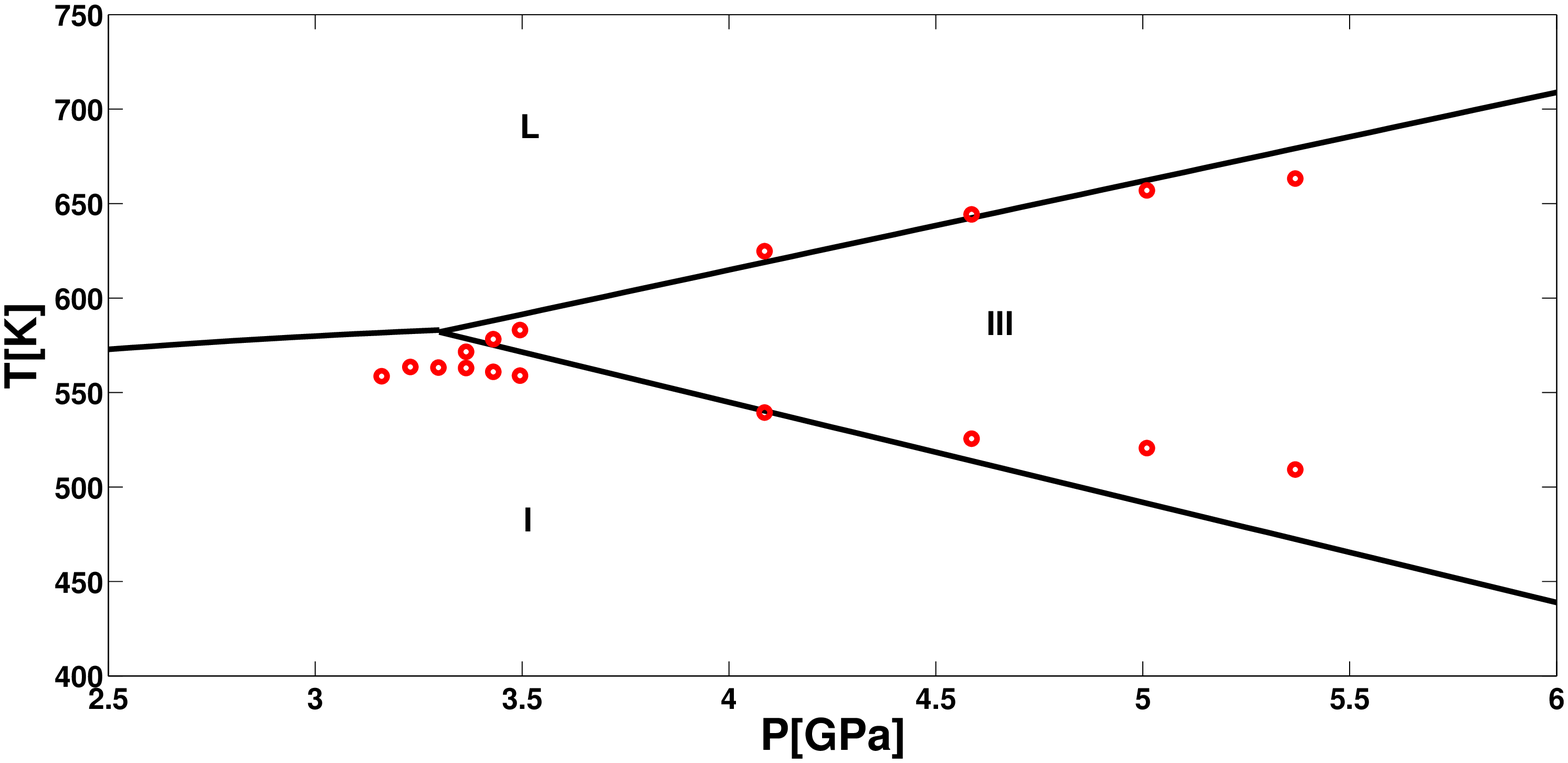}}
\subfigure[\label{fig:Phase Diagrams c}]{ \label{fig:Phase
Diagrams
c}\includegraphics[width=0.47\textwidth,height=0.25\textheight]{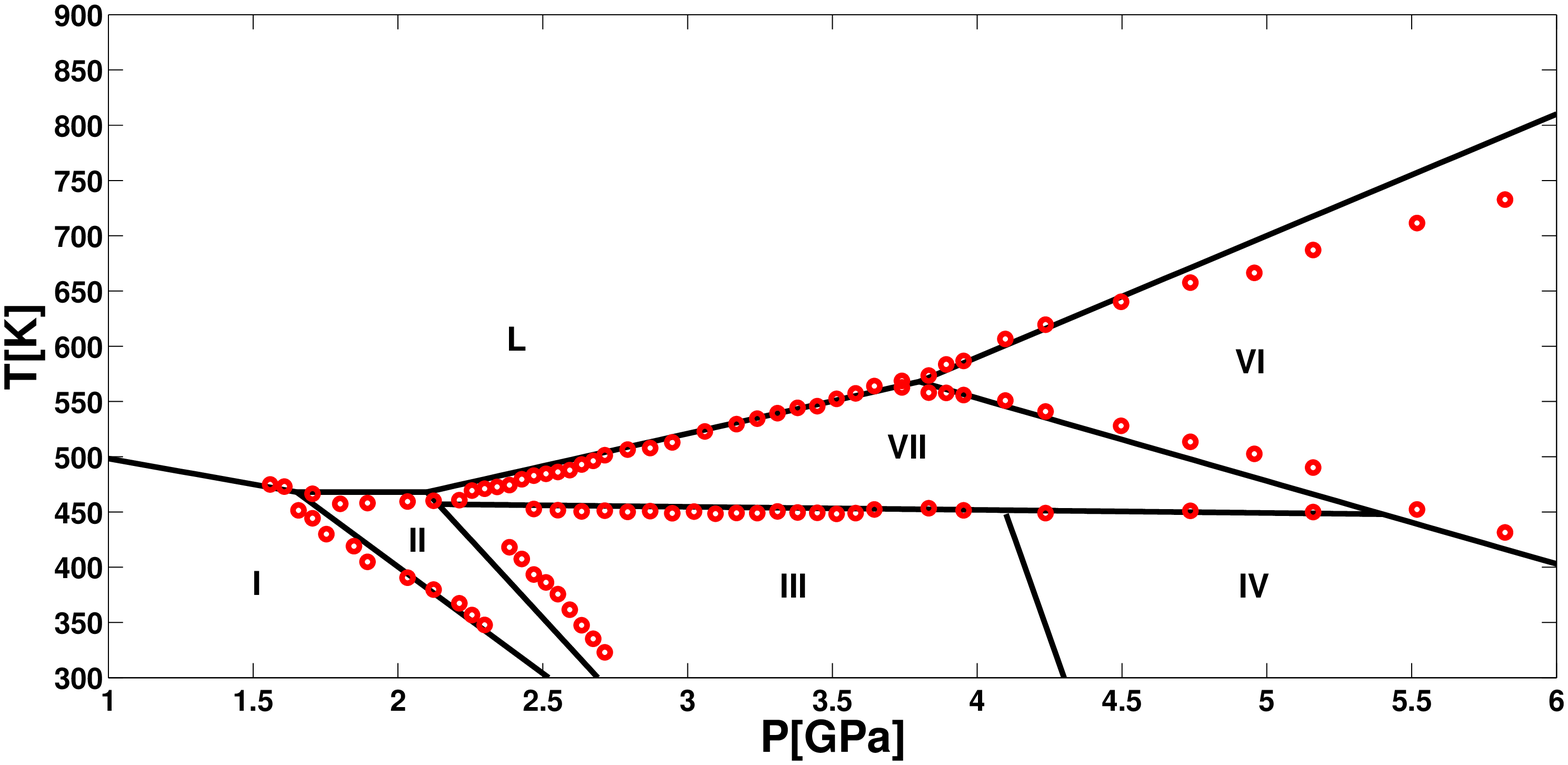}}
\end{center}
\caption{\label{fig:Phase Diagrams}(Color online) Comparison
between the phase diagrams of (a) Pb (b) Sn and (c) Bi derived
from our measurements (red points) and those previously published
in the literature (solid black
lines).\cite{Young,Tonkov&Ponyatovsky,DJD60,ED10,BFP58}}
\end{figure}

The quality of the calibration procedure and the data in Figs.\
\ref{fig:Pb_Isobars}-\ref{fig:Bi_Isobars} can be examined by
constructing the $P-T$ phase diagrams of Pb, Sn and Bi from our
electrical resistance measurements. In each of the isobars
presented in Figs.\ \ref{fig:Pb_Isobars}-\ref{fig:Bi_Isobars}, the
values of the oil pressure and heating power corresponding to the
onset of the phase transitions were identified. The corresponding
sample pressure and temperature were then derived using the
calibration presented above. The resulting phase diagrams are
shown in Fig.\ \ref{fig:Phase Diagrams}. The excellent agreement
between the phase diagrams derived from our data and those
reported in the literature is clearly
evident.\cite{Tonkov&Ponyatovsky,Young} This demonstrates that the
calibration does not vary within our experiments and well-defined
relations exist between the sample pressure and temperature and
the oil pressure and heating power within the pressure and
temperature range of the experiment (up to 6GPa and 1000K). These
relations are well modelled by Eqs.\ \eqref{eq:1} and
\eqref{eq:2}. Additionally, Fig.\ \ref{fig:Phase Diagrams}
supports the errors in estimating the sample pressure and
temperature indicated above, i.e.\ $\pm5\%$ in the sample pressure
and $\pm$20K in the sample temperature.
\section{Discussion and conclusions}
\label{Discussion} We have demonstrated a simple, low-cost and
sensitive experimental design for measurements of electrical
resistance at elevated pressures (up to 6GPa) and temperatures (up
to 1000K) using a PE large volume press. The sample is indirectly
heated by a resistive graphite heater. A detailed description of
the cell assembly is provided. Technical challenges associated
with the presented method, namely the disconnection of the
metallic electrodes during compression and the percolation of the
melt through the BN medium, were considered.

Measurements carried out on Pb, Sn and Bi samples are presented.
The measurements are of excellent quality, exhibiting clear
discontinuities in the electrical resistance of the sample at
phase transitions. These pronounced (nearly abrupt) changes
indicate that the pressure and temperature gradients across the
sample are very small. These measurements were used to calibrate
the sample pressure and temperature as a function of oil pressure
and heating power, as well as to validate the performance of the
experimental setup. It was found that a calibration independent of
sample type exists, thus making independent measurements of the
temperature and pressure redundant.

Based on the sample pressure and temperature calibration, the
$P-T$ phase diagrams of the above elements were reconstructed and
found to be in very good agreement with previously published ones.
We have thus shown that accurate electrical resistance
measurements can be carried out in a PE large volume press using
the indirect heating configuration and may be used to probe phase
transitions and anomalous behavior of condensed matter under
extreme conditions. Though entailing some technical challenges
compared to the direct heating configuration,\cite{PE08} the
quality of the data achieved in the indirect heating configuration
is much better. The small dimensions of the sample result in very
small pressure and temperature gradients. Furthermore, the
indirect heating configuration allows one to probe only the
sample, without the effect of the surrounding medium as in the
direct heating configuration.\cite{PE08}

Future experiments with the present method will be aimed at
investigations of elemental systems and alloys as well as
challenging studies of liquid-liquid phase transitions. In
particular, we believe that the high sensitivity of our
measurements, even to minor changes in electrical resistance, will
allow us to shed more light on liquid-liquid transitions.
\begin{acknowledgments}
We thank S. Klotz, Y. Le Godec, E. Sterer, R. Salem, M.
Nikolaevsky, A. Melchior and O. Noked for useful discussions. We
also thank A. Damri and E. Sterer for their help in producing cell
assembly components.
\end{acknowledgments}

\end{document}